# Magnetic and electronic properties of a topological nodal line semimetal candidate: HoSbTe

Meng Yang,[1,2,3,#] Yuting Qian,[1,3,#] Dayu Yan,[1,2,3] Yong Li,[1,2,3] Youting Song,[1] Zhijun Wang,[1,3] Changjiang Yi,[1] Hai L. Feng,[1] Hongming Weng,[1,3,4,*] and Youguo Shi,[1,2,3,*]

[1]*Beijing National Laboratory for Condensed Matter Physics and Institute of Physics, Chinese Academy of Sciences, Beijing 100190, China*

[2]*Center of Materials Science and Optoelectronics Engineering, University of Chinese Academy of Sciences, Beijing 100049, China*

[3]*School of Physical Sciences, University of Chinese Academy of Sciences, Beijing 100190, China*

[4]*Songshan Lake Materials Laboratory, Dongguan, Guangdong 523808, China*

[#]These authors contributed equally to this work.

[*]Corresponding authors: hmweng@iphy.ac.cn (HW); ygshi@iphy.ac.cn (YS)

# Abstract

We report the experimental and theoretical studies of a magnetic topological nodal line semimetal candidate HoSbTe. Single crystals of HoSbTe are grown from Sb flux, crystallizing in a tetragonal layered structure (space group: *P4/nmm*, no. 129), in which the Ho-Te bilayer is separated by the square-net Sb layer. The magnetization and specific heat present distinct anomalies at ~ 4 K related to an antiferromagnetic (AFM) phase transition. Meanwhile, with applying magnetic field perpendicular and parallel to the crystallographic *c* axis, an obvious magnetic anisotropy is observed. Electrical resistivity undergoes a bad-metal-like state below 200 K and reveals a plateau at about 8 K followed by a drop due to the AFM transition. In addition, with the first-principle calculations of band structure, we find that HoSbTe is a topological nodal line semimetal or a weak topological insulator with or without taking the spin-orbit coupling (SOC) into account, providing a new platform to investigate the interplay between magnetic and topological fermionic properties.

# Introduction

Topological nodal line semimetals (NLSMs) are defined as a system where the conduction and valance bands cross with each other near the Fermi level and the crossing points form a closed loop or periodically continuous line in momentum space [1-8]. Compared with Dirac/Weyl semimetals, NLSMs exhibit more complex topological configurations, such as nodal ring, nodal chain [9], nodal link, and nodal knot [10], which are expected to lead to many quantum phenomena including a fat Landau level, long-range Coulomb interaction, the Kondo effect, and a quasi-topological electromagnetic response [11-14]. Numerous NLSMs have been theoretically predicted, such as CaAgX (X = P, As) [4], carbon allotropes [6], black phosphorus under pressure [7], antiperovskite $Cu_3PdN$ [8], and so on. However, only a few candidates have been verified by experiments [15,16]. Searching for new NLSMs is of great significance.

Recently, the *WHM* compounds (*W* = Zr, Hf, or La, *H* = Si, Ge, Sn, or Sb, and *M* = O, S, Se, or Te), predicted as stacking of two-dimensional (2D) topological insulators (TIs) by Q. Xu *et al* when spin-orbit coupling (SOC) is considered and as NLSMs when SOC is neglected, have attracted considerable attention [17]. ZrSnTe, one of the *WHM* family, hosts 2D electronic bands of TI state, attested by using angle-resolved photoemission spectroscope (ARPES) combined with the first-principle calculations [18]. The other member ZrSiS has been experimentally confirmed as a topological NLSM by ARPES and transport measurements [19-21]. Also, nodal line fermions in ZrSiSe and ZrSiTe have been proved by ARPES [22]. Recently, LaSbTe is expected to be a weak topological insulator when SOC is taken into account, and a crystal structural phase transition leads it to topological crystalline insulator [23]. Substituting La with other rare-earth elements with non-zero *4f* electrons can introduce the magnetic couplings and correlated electronic effect, and new physical phenomena will emerge due to the interplay between magnetism and topological bands. One example is the AFM CeSbTe (with one *4f* electron) which shows magnetically tunable Weyl and Dirac states or an eightfold band cross at a high-symmetry point [24]. The high-order band degeneracies in CeSbTe may be created by the AFM order. Another example, AFM GdSbTe, is a topological nodal line semimetal demonstrated by ARPES [25]. These studies indicate that LnSbTe (Ln = rare-earth elements with non-zero 4*f* electrons) is a promising platform to search for new exotic topological electronic states.

In this work, single crystals of HoSbTe, a new member of the *WHM* family, are successfully grown. HoSbTe is a bad-metal and shows an AFM order below 4 K. In the absence of SOC, the first-principle band calculations indicate that HoSbTe is an ideal NLSM and all band crosses are close to the Fermi level. When SOC is considered, all of the band crosses will open a small gap. Our discovery shall give access to enrich the *WHM* family and expand the magnetic NLSMs.

## Experiment detail

High-quality single crystals of HoSbTe were grown by a Sb-flux method. Starting materials of Ho, Sb, and Te with a mole ratio of 1: 20 :1 were mixed and placed into an alumina crucible. The crucible was then placed into a quartz tube and sealed under a high vacuum atmosphere. The ampoule was heated to 1373 K over 15 h and maintained for 10 h to ensure homogeneous mixing of the starting materials. Then it was slowly cooled to 1073 K at a rate of 2 K/h, followed by centrifuging to separate the crystals from the excess Sb. Shine-sheet crystals with metallic luster were obtained, as shown in the inset of Fig. 1(b).

Single-crystal x-ray diffraction (XRD) measurement was conducted on Bruker D8 Venture High-Resolution Four-Circle Diffractometer at 273 K using Mo $K_\alpha$ radiation ($\lambda$ = 0.71073 Å). The data were refined by a full-matrix least-squares of $F^2$ via the SHELXL-2016/6 program [26]. The *(001)* surface of the HoSbTe single crystal was studied via a Bruker D8 Advance XRD detector by using Cu $K_{\alpha 1}$ radiation ($\lambda$ = 1.5418 Å). The analysis of chemical composition was carried out using energy-dispersive x-ray spectroscopy installed on a Hitachi S-4800 scanning electron microscope at an accelerating voltage of 15 kV. Magnetizations (*M*) were measured between 2 K to 300 K with varied applied fields ($H$ = 1, 10, 30, 50 kOe) in the field-cooling (FC) and zero-field-cooling (ZFC) modes using a Magnetic Properties Measurement System (MPMS, Quantum Design Inc.) with a SQUID-VSM option. Isothermal magnetizations $M(H)$ were measured at 2 K and 300 K with applied fields between -70 kOe to 70 kOe. During the measurements, the $c$ axis of the single crystal was set parallel and perpendicular to the external magnetic field, respectively. The heat capacity ($C_P$) was measured from 2 K to 300 K without/with applied magnetic fields in a Physical Property Measurement System (PPMS, Quantum Design Inc.). The longitudinal resistivity was measured in the same PPMS with a configuration of four wires by Indium contact.

Theoretical calculations were performed using the full-potential linearized-augmented plane wave (FP-LAPW) method implemented in the WIEN2K software package [27]. The exchange-correlation of Perdew, Burke and Ernzerhof within generalized gradient approximation (GGA) was applied in the calculations [28]. The atomic sphere radius RMT for Ho, Sb, and Te were 2.5 bohrs. The truncation of the modulus of the reciprocal lattice vector Kmax was set to RMTKmax = 7. The $k$-mesh grid of the BZ was 8×8×3. The GGA + Hubbard-U (GGA + U) method was adopted in the FM and AFM calculations, and the exchange parameter $U$ = 7 eV was applied to the Ho 4$f$ states. The effect of SOC was included in a second-variational procedure. The crystal structure from our single-crystal XRD study was exploited. The electronic structures were carried out both without and with considering the SOC.

## Result and discussion

Single-crystal XRD data of HoSbTe were well refined with a tetragonal space group $P4/nmm$ (no. 129) which was also adapted by sister compounds CeSbTe and GdSbTe. The crystallographic data for HoSbTe are summarized in Table I and the lattice parameters are $a$ = $b$ = 4.2294(3) Å and $c$ = 9.1458(9) Å, which are both smaller than CeSbTe and GdSbTe [24, 29]. In this space group, the Sb occupies the Wyckoff position 2b (1/4,3/4,1/2) while Ho and Te each occupy a 2c (3/4,3/4,$z$) site. During the refinement, occupancies on each site were examined and there was no indication of deficiency. The occupancies were calculated to be 1 in the final refinement. The refined atomic positions are summarized in Table II.

Table I Crystallographic and structure refinement data for HoSbTe.

| | |
|---|---|
| Empirical formula | HoSbTe |
| Formula weight (g mol$^{-1}$) | 414.28 |
| Temperature | 273(2) K |
| Wavelength | Mo $K\alpha$ (0.71073 Å) |
| Crystal system | tetragonal |
| Space group | *P4/nmm* |
| Unit cell dimensions | $a = b = 4.2294(3)$ Å, $c = 9.1458(9)$ Å |
| Cell volume (Å) | 163.60(3) |
| Z | 2 |
| Density (g cm$^{-3}$) | 8.410 |
| *F*(000) | 340.0 |
| *h k l* range | $-5 \leq h \leq 5$, $-5 \leq k \leq 4$, $-11 \leq l \leq 11$ |
| $\theta_{min}$ (°), $\theta_{max}$ (°) | 2.227, 28.227 |
| Linear absorption coeff. (mm$^{-1}$) | 40.685 |
| No. of reflections | 1069 |
| No. independent reflections | 151 |
| No. observed reflections | 144 |
| R factors | 2.85% ($R_1[F_0 > 4\sigma(F_0)]$), 5.73% (w$R_2$) |
| Weighting scheme | w = $1/[\sigma^2(F_o^2) + (0.0196P)^2 + 0.1805P]$ where $P = (F_o^2 + 2F_c^2)/3$ |
| Refinement software | SHELXL-2016/6 |

Table II Atomic coordinates and equivalent isotropic thermal parameters of HoSbTe.

| Site | Wyckoff | x | y | z | Occupation | $U_{eq}$ (Å$^2$) |
|---|---|---|---|---|---|---|
| Ho | 2*c* | 0.75 | 0.75 | 0.77647(1) | 1 | 0.0067(4) |
| Sb | 2*b* | 0.25 | 0.75 | 0.5 | 1 | 0.0072(4) |
| Te | 2*c* | 0.25 | 0.25 | 0.87579(14) | 1 | 0.0063(4) |

The refined crystal structure of HoSbTe is displayed in Fig. 1(a) and the Ho-Te bilayers are sandwiched between the square-net Sb layers, which is similar to that series of ZrSiX (X = S, Se, Te) [30, 22] and LnSbTe (Ln = La, Ce, Gd) [23, 24, 29]. The interlayer Ho-Sb bond length is 3.2963(8) Å × 4. The interlayer Ho-Te bond length is 3.1804(2) Å × 1 which is longer than the intralayer Ho-Te bond (3.1255(5) Å × 4). These bond lengths are slightly shorter than those in GdSbTe [29] (intralayer Gd-Te =3.17 Å × 4; interlayer Gd-Te = 3.21 Å × 1), which is reasonable because that the ionic radius of $Ho^{3+}$ is somewhat shorter than $Gd^{3+}$ [31]. The XRD patterns on a flat surface of HoSbTe single crystal are displayed in Fig. 1(b) and shows only *(00l)* peaks, revealing high crystalline quality.

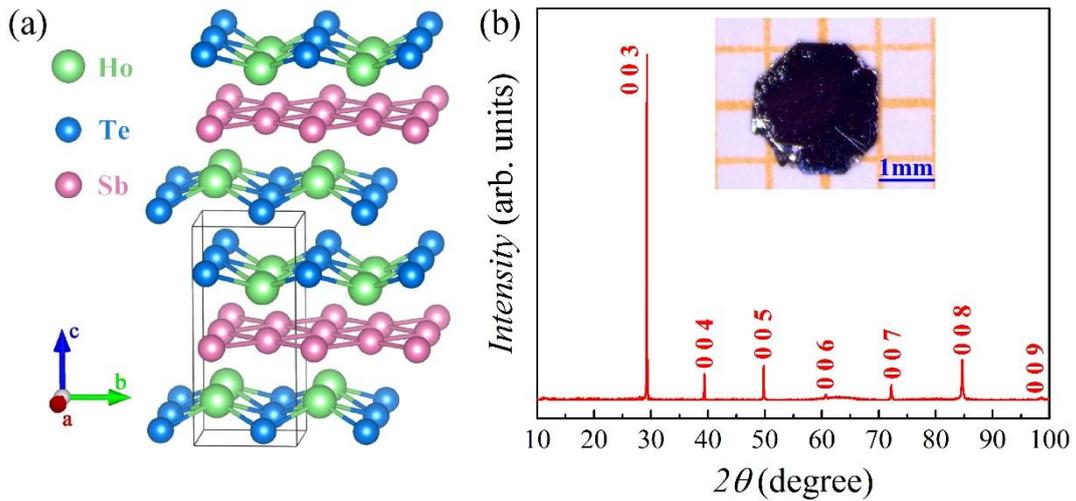

FIG. 1. (a) Crystal structure of HoSbTe. Green, blue and pink balls represent the Ho, Te and Sb atoms, respectively. (b) The XRD pattern for a HoSbTe single crystal and only a series of *(00l)* peaks were shown. The inset is a photograph of a typical single crystal (the back square is 1 × 1 $mm^2$).

Temperature dependent magnetization curves, *M(T)*, measured with various magnetic fields for *H* = 1, 10, 30, 50 kOe are shown in Fig. 2(a) (*H* ⊥ *c*) and Fig. 2(b) (*H* // *c*). The results of the FC mode are concealed because of the superimposition with the ZFC data. When *H* ⊥ *c*, the *H* = 1 kOe curve shows a sharp peak at ~ 4 K, indicating an AFM transition. But the peak disappears when the field is up to 10 kOe or larger (see the inset of Fig. 2(a)), implying a possible field-induced spin reorientation. The magnetizations at low temperatures are over 11 $\mu_B$/f. u. for $H = 50$ kOe, suggesting a fully polarized ferromagnetic (FM) state. For $H // c$, all the curves show cusps at ~ 4 K, indicating a weak magnetic dependence for AFM order. The different magnetic behavior between $H \perp c$ and $H // c$ indicates a strong magnetic anisotropy in HoSbTe.

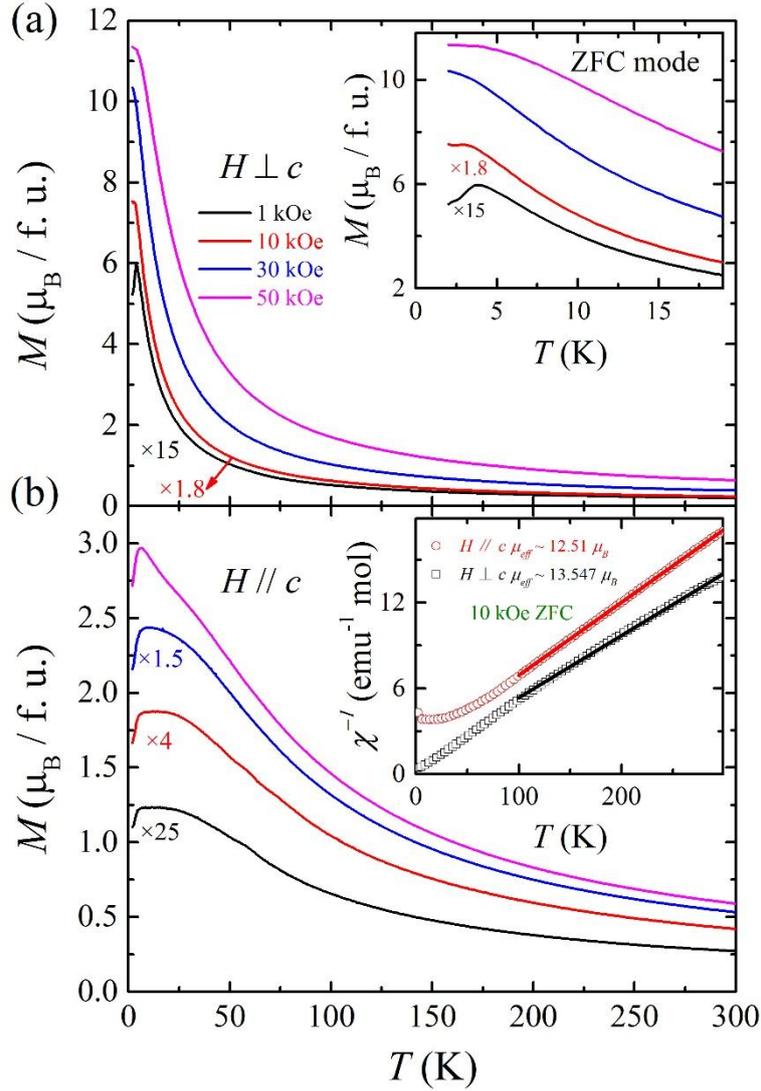

FIG. 2. Temperature dependence of the ZFC magnetization $M$ (scaled by different factors) with $H = 1, 10, 30, 50$ kOe, measured in perpendicular (a) and parallel (b) direction of $c$ axis to the field. The inset of in Fig. 2(a) is a close-up view of magnetization curves at low temperature for $H \perp c$. The inset of Fig. 2(b) shows the Curie-Weiss fit (solid lines) to the ZFC inverse susceptibility in a field with $H = 10$ kOe above 100 K.

The inset of Fig. 2(b) presents the temperature dependence of the ZFC inverse susceptibility ($\chi^{-1}$) with $H$ = 10 kOe. For both $H \perp c$ and $H // c$ cases, the data above 100 K can be well fitted by the Curie-Weiss law. The analytical formula is $\chi^{-1} = T/C - \theta_W/C$, where $C$ and $\theta_W$ are the Curie constant and the Weiss temperature, respectively. From the fit, we obtain negative $\theta_W$ of -22.5 K (-35.2 K) for $H \perp(//) c$, revealing strong antiferromagnetic interactions between the magnetic moments of Ho ions. The effective magnetic moments $\mu_{eff}$ are estimated to be 13.547 $\mu_B$ (12.51 $\mu_B$) for $H \perp(//) c$, which are not far from the theoretical value of 10.6 $\mu_B$ for a $Ho^{3+}$ ion. The relative larger experimental $\mu_{eff}$ may be caused by the crystal electric field created by the surrounding ligands of $Ho^{3+}$ ions, as mentioned in other Ho compounds [32-36].

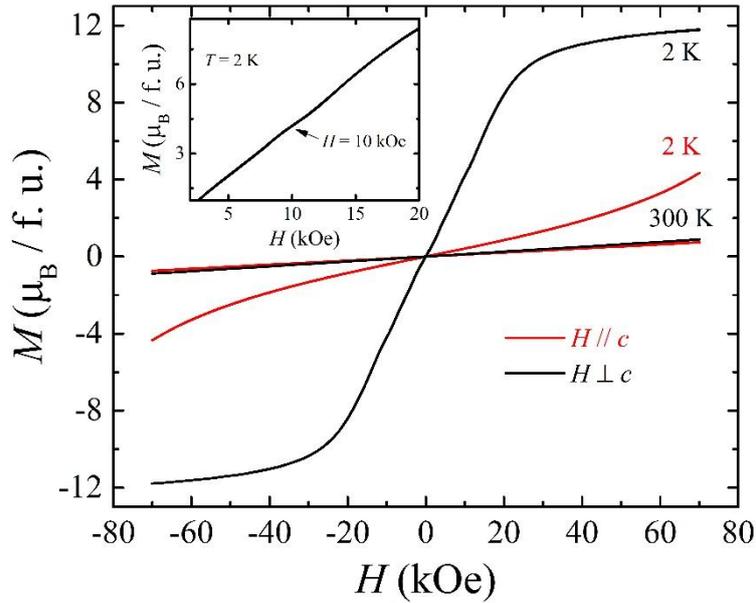

FIG. 3. Magnetization $M$ plotted as a function of magnetic-field $H$ for $H // c$ and $H \perp c$ axis at 2 K and 300 K. The inset shows the details at 2 K for $H \perp c$.

To further probe the magnetic order, isothermal magnetizations measured (at 2 and 300 K) below and above the transition temperature were shown in Fig. 3. At 300 K, both curves show linear behavior and small magnetization values, which is consistent with the paramagnetic (PM) feature. An obvious anisotropic magnetization is observed at 2 K. For $H \perp c$, the magnetization increases rapidly and almost saturates at 70 kOe. However, the magnetization grows monotonically and is far from saturation when $H //$

*c*. The magnetic moment at 2 K and 70 kOe for $H \perp c$ case is approximately 11.76 $\mu_B$. Meanwhile, a flexure was observed at around 10 kOe at 2 K, as seen in the inset of Fig. 3, indicating a possible spin-flop transition. For $H // c$, the magnetization is far from saturation, but the *M(H)* curves are not linear and show an upturn when fields are larger than 40 kOe, indicating a possible spin-flop transition if higher fields would be applied. Spin-flop transitions and magnetic anisotropy were also reported in similar compounds CeSbSe, CeSbTe and GdSbTe [37, 24, 25].

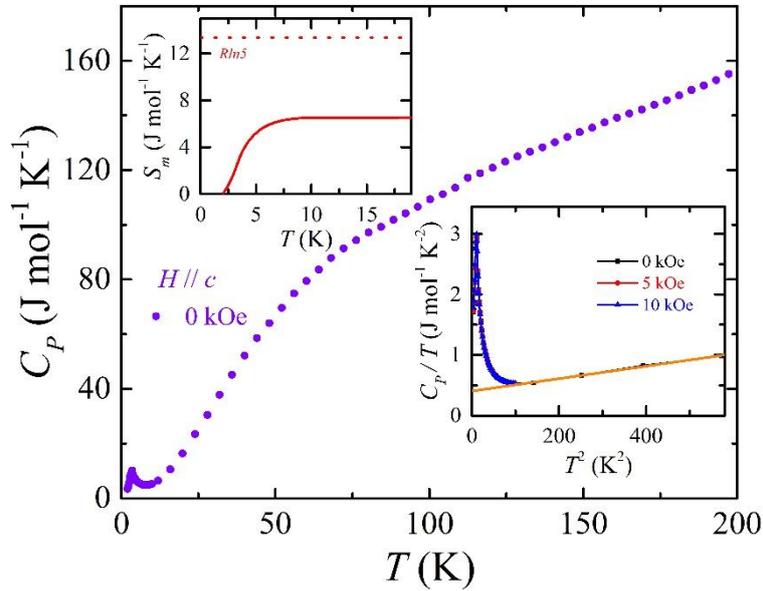

FIG. 4. The specific heat capacity measured without applied fields for $H // c$. The right inset shows the $C_P/T$-$T^2$ plots at low temperature with $H$ = 0, 5, 10 kOe. The orange solid line is a fit to the data using the Fermi-liquid model as described in the text. The left inset shows the magnetic entropy change as a function of temperature.

The specific heat capacity ($C_P$) of HoSbTe with zero applied magnetic field was shown in Fig. 4. The $C_P$ decreases monotonically down to 10 K and then a λ-shape peak near 4 K is observed, which confirms the long-range AFM order observed in magnetic properties. Similar $C_P$ anomalies were observed in other AFM isostructure materials, such as CeSbSe, CeSbTe, GdSbTe [37, 24, 25]. The right inset in Fig. 4 shows the low temperature $C_P / T$ vs. $T^2$ plots measured with $H$ = 0, 5, and 10 kOe ($H // c$). The sharp peak is immune to the magnetic fields, indicating a robust AFM order, which is in accordance with the $M(T)$ data for $H // c$. A measurement under a higher magnetic field was failed because the sample was pulled away from the stage of the heat capacity puck when the magnetic fields were up to 30 kOe.

The $C_P/T$ vs. $T^2$ data from 10 K to 20 K was fitted by the Fermi-liquid model using the expression for $C/T = \gamma + \beta T^2$, where $\gamma$ is the electronic coefficient of the heat capacity and $\beta T^2$ is the low temperature limit of the lattice term [37]. The fit yields $\beta = $ 1.12 mJ/mol/K$^4$ and $\gamma = $ 382.2 mJ/mol/K$^2$, which is unexpectedly high and almost comparable to some heavy fermion materials. Based on this fit, we obtain the magnetic entropy $S_m$ by subtracting the contribution of electron and lattice terms (see the left inset of Fig. 4). The $S_m$ rapidly increases and reaches saturation of ~ 6.52 J/mol/K, which is nearly half of the theoretical value of 13.374 J mol$^{-1}$ K$^{-1}$ calculated from $S_{cal} = R\ln(2S+1)$ for spin-only configuration. The relatively small $S_m$ may indicate that the lattice contribution was overestimated. It is noteworthy that a non-magnetic reference compound may be better as the background of electron and lattice contribution. Unfortunately, the LuSbTe single crystals are difficult to synthesis.

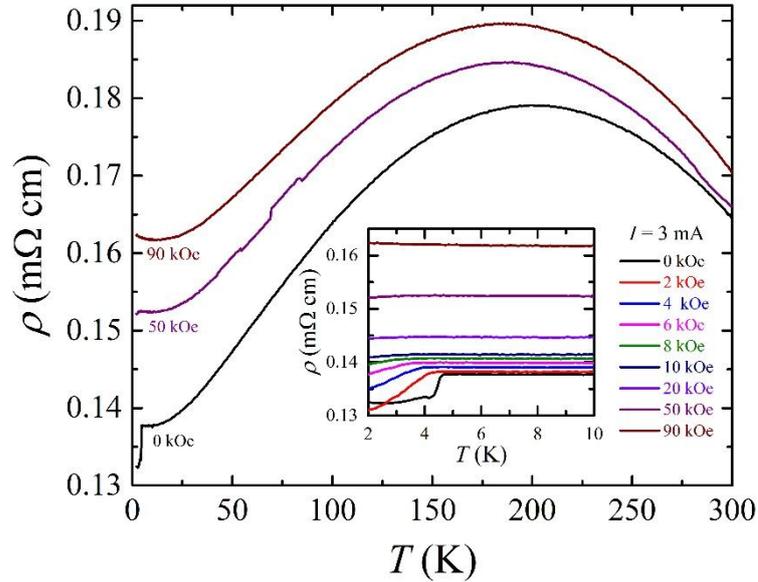

FIG. 5. Resistivity $\rho$ plotted as a function of temperature $T$ at various applied fields for $H // c$. The inset is an enlarged view below 10 K.

Temperature dependence of resistivity $\rho(T)$ curves measured at various applied fields for *H // c* are shown in Fig. 5. On cooling, ρ increases first and shows a broad hump-like feature at ~ 200 K, where the sample is in the PM phase. Hence, the origin of this hump may need some special measurements to clarify. On further cooling *T*, ρ decreases until a drop appears near 4 K, supporting the consideration of AFM order. When magnetic fields are applied, ρ increases with increasing magnetic fields. An enlarged view of ρ below 10 K is shown as the inset of Fig. 5. The drop slightly shifts to lower temperatures with increasing of *H*, and is completely suppressed when *H* is larger than 10 kOe. As a plateau of ρ occurs at around 8 K, the residual resistivity ratio defined as [*RRR* = ρ(200 K)/ρ(8 K)] can be roughly calculated to be ~ 1.301, indicating a bad-metal-like state in HoSbTe.

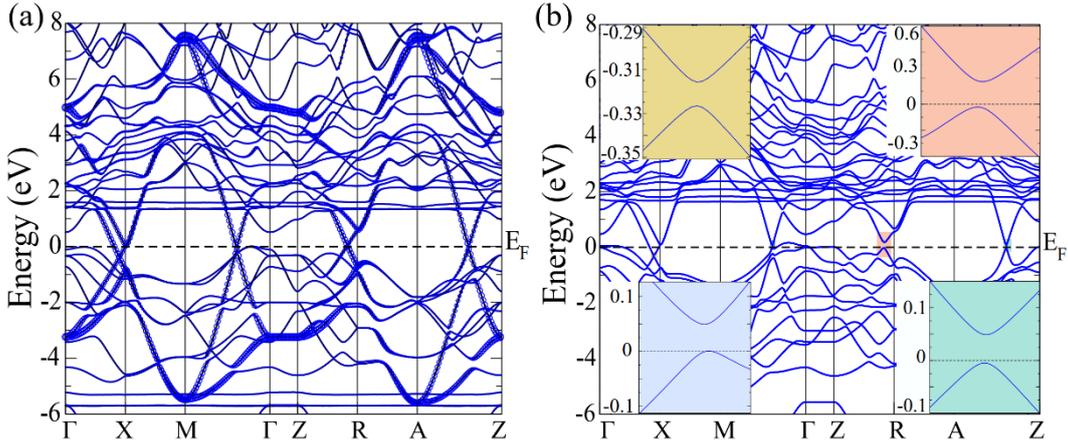

FIG. 6. The calculated band structure for AFM HoSbTe (a) without SOC and (b) with SOC. (a)The size of the blue circle corresponds to the weight of $p_x$ and $p_y$ orbits of Sb. (b) The insets are the zoom in of the anti-crossing points.

By calculating total energy, the AFM is more stable than FM state in HoSbTe about 2 meV/f.u., which is consistent with the AFM state we observed from the magnetization curves around 4 K. We assume the magnetic moment is along the *(001)* direction and the spin ordering is A-type, i.e., the intra-plane coupling is ferromagnetic while inter-plane coupling is antiferromagnetic. Thus, the magnetic space group (MSG) is 129.419 [38]. The calculated band structures for AFM HoSbTe without and with SOC are shown in Fig. 6(a) and 6(b), respectively. In Fig. 6(a), the size of the blue circle is proportional to the weight of $p_x$, $p_y$ orbitals in Sb atoms. Since the 4*f* electrons of Ho mainly contribute to local magnetic moments and they are far away from the Fermi level, the essential characteristics of the band topology can be completely captured by the Sb atomic layer similar with the properties of LaSbTe and other *WHM* family compounds [17,23]. If SOC is not taken into consideration, there are nodal points along Γ-M, Γ-X(Y) at $k_z = 0$ and Z-R, Z-A at $k_z = \pi$ planes protected by glide mirror symmetry perpendicular to the *(001)* surface, which is actually a part of nodal circles. When SOC is considered, time-reversal symmetry, space inversion symmetry and all the mirror symmetries are broken in this AFM magnetic structure. There is a full band gap with SOC and there is no symmetry-based indicator that can be defined in this MSG [39]. There is no band inversion along Γ-Z, which is a very typical two-dimensional (2D) feature [17]. The topological feature of AFM HoSbTe can be understood as following. In each spin channel, each layer is a 2D Chern insulator and stack weakly along *c* direction [23,40]. The two opposite spin channels are degenerate in the ideal AFM case. If AFM is ignored, the fully gapped nonmagnetic bulk state would be a weak TI with symmetry-based indicator $Z_{2,2,2,4}$ = (0010).

## Conclusion

Combining the measurements of physical properties on single crystals and first-principle band calculations, we present a systematic study of a magnetic topological semimetal candidate HoSbTe. As a new member of the *WHM* materials, HoSbTe crystallizes in a tetragonal layered structure with the Ho-Te bilayer and Sb layer being alternately stacked. A PM-AFM phase transition was observed at ~ 4 K, indicating that the AFM interaction between $Ho^{3+}$ ions dominants the magnetic ground states. Meanwhile, HoSbTe exhibits a large anisotropic magnetic property at low temperature with the magnetization of $H \perp c$ saturating faster than $H // c$. Electrical resistivity measurements indicate an unexpected bad-metal-like dependence with temperature, which is not as similar as the phenomenon observed in other *WHM* materials. However, our first-principle calculations of band structure propose that the topological NLSM or weak TI state would emerge without or with considering the effect of SOC in the AFM

HoSbTe, which makes it a promising candidate for exploring magnetic topological materials. Although the results of our experiments are rather limited to notarize the existence of topological NLSM state in HoSbTe, further measurements on ARPES shall bring forward the evidence of the existence of topological band structure.

## Acknowledgments

We thank Gang Li and Zhenyu Mi for valuable discussions. This work was supported by the National Key Research and Development Program of China (Grants No. 2017YFA0302901, 2016YFA0300604), the National Natural Science Foundation of China (Grants No. 11774399, No. 11674369, No. 11925408, No. 11921004, and No. 11974395 ), Beijing Natural Science Foundation (Grant No. Z180008), the Beijing Municipal Science and Technology Commission (Z181100004218001), the Strategic Priority Research Program of Chinese Academy of Sciences (Grant No. XDB33000000), the K. C. Wong Education Foundation (Grant No. GJTD-2018-01), and the CAS Pioneer Hundred Talents Program.

## Reference

1. H. M. Weng, X. Dai, and Z. Fang, Topological semimetals predicted from first-principles calculations, *J. Phys.: Condens. Matter* 28 303001 (27pp) (2016)
2. C. Fang, H. M. Weng, X. Dai, and Z. Fang, Topological nodal line semimetals, *Chin. Phys.* B 25, 117106 (2016)
3. A. A. Burkov, M. D. Hook, and Leon Balents, Topological nodal semimetals, *Phys. Rev.* B 84, 235126 (2011)
4. A. Yamakage, Y. Yamakawa, Y. Tanaka, and Y. Okamoto, Line-Node Dirac Semimetal and Topological Insulating Phase in Noncentrosymmetric Pnictides CaAgX (X = P, As), *J. Phys. Soc. Japan* 85 013708 (2016)
5. Q. F. Liang, J. Zhou, R. Yu, Z. Wang, and H. M. Weng, Node-surface and node-line fermions from nonsymmorphic lattice symmetries, *Phys. Rev.* B 93, 085427 (2016)
6. H. M. Weng, Y. Y. Liang, Q. N. Xu, R.Yu, Z. Fang, X. Dai, and Y. Kawazoe, Topological node-line semimetal in three-dimensional graphene networks, *Phys. Rev.* B 92, 045108 (2015)
7. J. Z Zhao, R. Yu, H. M. Weng, and Z. Fang, Topological node-line semimetal in

compressed black phosphorus, *Phys. Rev.* B 94, 195104 (2016)

8. R. Yu, H. M. Weng, Z. Fang, X. Dai, and X. Hu, Topological Node-Line Semimetal and Dirac Semimetal State in Antiperovskite $Cu_3PdN$, *Phys. Rev. Lett.* 115, 036807 (2015)

9. T. Bzdušek, Q. S. Wu, A. Rüegg, M. Sigrist, and A. A. Soluyanov, Nodal-chain metals, *Nature* 538, 75 (2016)

10. R. Bi, Z. B. Yan, L. Lu, and Z. Wang, Nodal-knot semimetals, *Phys. Rev.* B 96, 201305 (2017)

11. J.-W. Rhim and Y. B. Kim, Landau level quantization and almost flat modes in three-dimensional semimetals with nodal ring spectra, *Phys. Rev.* B 92, 045126 (2015)

12. Y. J. Huh, E.-G. Moon, and Y. B. Kim, Long-range Coulomb interaction in nodal-ring semimetals, *Phys. Rev.* B 93, 035138 (2016)

13. A. K. Mitchell and L. Fritz, Kondo effect in three-dimensional Dirac and Weyl systems, *Phys. Rev.* B 92, 121109(R) (2015)

14. S. T. Ramamurthy and T. L. Hughes, Quasitopological electromagnetic response of line-node semimetals, *Phys. Rev.* B 95, 075138 (2017)

15. G. Bian, T.-R. Chang, R. Sankar, S.-Y. Xu, H. Zheng, T. Neupert, C.-K. Chiu, S.-M. Huang, G. Chang, I. Belopolski, D. S. Sanchez, M. Neupane, N. Alidoust, C. Liu, B. Wang, C.-C. Lee, H.-T. Jeng, C. Zhang, Z. Yuan, S. Jia, A. Bansil, F. Chou, H. Lin, and M. Z. Hasan, Topological nodal-line fermions in spin-orbit metal $PbTaSe_2$, *Nat. Commun.* 7, 10556 (2016).

16. C.-J. Yi, B. Q. Lv, Q. S. Wu, B.-B. Fu, X. Gao, M. Yang, X.-L. Peng, M. Li, Y.-B. Huang, P. Richard, M. Shi, G. Li, O. V. Yazyev, Y.-G. Shi, T. Qian, and H. Ding, Observation of a nodal chain with Dirac surface states in $TiB_2$. *Phys. Rev.* B 97, 201107(R) (2018)

17. Q. N. Xu, Z. D. Song, S. M. Nie, H. M. Weng, Z. Fang, and X. Dai, Two-dimensional oxide topological insulator with iron-pnictide superconductor LiFeAs structure, *Phys. Rev.* B 92, 205310 (2015)

18. R. Lou, J.-Z. Ma, Q.-N. Xu, B.-B. Fu, L.-Y. Kong, Y.-G. Shi, P. Richard, H.-M. Weng, Z. Fang, S.-S. Sun, Q. Wang, H.-C. Lei, T. Qian, H. Ding, and S.-C. Wang, Emergence of topological bands on the surface of ZrSnTe crystal, *Phys. Rev.* B 93, 241104(R) (2016)

19. B.-B. Fu, C.-J. Yi, T.-T. Zhang, M. Caputo, J.-Z. Ma, X. Gao, B. Q. Lv, L.-Y. Kong, Y.-B. Huang, P. Richard, M. Shi, V. N. Strocov, C. Fang, H.-M. Weng, Y.-G. Shi, T. Qian, and H. Ding, Dirac nodal surfaces and nodal lines in ZrSiS, *Sci. Adv.* 5 eaau6459 (2019)

20. M. N. Ali, L. M. Schoop, C. Garg, J. M. Lippmann, E. Lara, B. Lotsch, and S. S. P. Parkin, Butterfly magnetoresistance, quasi-2D Dirac Fermi surface and topological

phase transition in ZrSiS, *Sci. Adv.* 2 e1601742 (2016)

21. R. Singha, A. K. Pariari, B. Satpati, and P. Mandal, Large nonsaturating magnetoresistance and signature of nondegenerate Dirac nodes in ZrSiS, *Proc. Natl. Acad. Sci.* USA 114, 2468 (2017)
22. J. Hu, Z. J. Tang, J. Y. Liu, X. Liu, Y. L. Zhu, D. Graf, K. Myhro, S. Tran, C. N. Lau, J. Wei, and Z. Q. Mao, Evidence of Topological Nodal-Line Fermions in ZrSiSe and ZrSiTe, *Phys. Rev. Lett.* 117, 016602 (2016)
23. Y. T. Qian, Z. Y. Tan, T. Zhang, J. C. Gao, Z. J. Wang, Z. Fang,1,2 Chen Fang, and H. M. Weng, Layer Construction of Topological Crystalline Insulator LaSbTe, *Sci. China Phys. Mech.* 63, 107011 (2020)
24. L. M. Schoop, A. Topp, J. Lippmann, F. Orlandi, L. Müchler, M. G. Vergniory, Y. Sun, A. W. Rost, V. Duppel, M. Krivenkov, S. Sheoran, P. Manuel, A. Varykhalov, Bc. H. Yan, R. K. Kremer, C. R. Ast, and B. V. Lotsch, Tunable Weyl and Dirac states in the nonsymmorphic compound CeSbTe, *Sci. Adv.* 4 eaar2317 (2018)
25. M. M. Hosen, G. Dhakal, K. Dimitri, P. Maldonado, A. Aperis, F. Kabir, C. Sims, P. Riseborough, P. M. Oppeneer, D. Kaczorowski, T. Durakiewicz, and M. Neupane, Discovery of topological nodalline fermionic phase in a magnetic material GdSbTe, *Sci. Rep.* 8, 13283(2018)
26. G. M. Sheldrick and T. R. Schneider, 1997 SHELXL: Highresolution refinement, *Methods Enzymol.* 277, 319 (1997)
27. P. Blaha, K. Schwarz, G. K. H. Madsen, D. Kvasnicka, J. Luitz, R. Laskowski, F. Tran, and L. D. Marks, WIEN2K, An Augmented Plane Wave + Local Orbitals Program for Calculating Crystal Properties (Karlheinz Schwarz, Techn. Universität Wien, Austria, 2001).
28. J. P. Perdew, K. Burke, and M. Ernzerhof, Generalized Gradient Approximation Made Simple, *Phys. Rev. Lett.* 77, 3865-3868 (1996)
29. R. Sankar, I. P. Muthuselvam, K. R. Babu, G. S. Murugan, K. Rajagopal, R. Kumar, T. C. Wu, C. Y. Wen, W. L. Lee, G. Y. Guo, and F. C. Chou, Crystal Growth and Magnetic Properties of Topological Nodal-Line Semimetal GdSbTe with Antiferromagnetic Spin Ordering, *Inorg. Chem.* 58, 17, 11730 (2019)
30. L. M. Schoop, M. N. Ali, C. Straßer, A. Topp, A. Varykhalov, D. Marchenko, V. Duppel, S. S. P. Parkin, B. V. Lotsch, and C. R. Ast, Dirac cone protected by non-symmorphic symmetry and three-dimensional Dirac line node in ZrSiS, *Nat. Commun.* 7, 11696 (2016)
31. R. D. Shannon, Revised Effective Ionic Radii and Systematic Studies of Interatomic Distances in Halides and Chalcogenides, *Acta Cryst.* A 32, 751-67 (1976)
32. J. Albino Aguiar, D. A. Landinez Tellez, Y. P. Yadava, and J. M. Ferreira, Structural and magnetic properties of the complex perovskite oxide $Ba_2HoHfO_{5.5}$, *Phys. Rev.* B 58, 2454 (1998)

33. S. Skanthakumar, C. K. Loong, and L. Soderholm, Crystal-field excitations and magnetic properties of Ho$^{3+}$ in HoVO$_4$, *Phys. Rev.* B 51, 12451 (1995)
34. I. L. Sashin, E. A. Goremychkin, A. Szytula, and E. S. Clementyev, Crystal Electric Field in RAgSb$_2$ (R = Ho, Er, Tm) Intermetallic Compounds, *Crystallogr. Rep.* 52, 412-9 (2007)
35. B. Z. Malkin, T. T. A Lummen, P. H. M. van Loosdrecht, G. Dhalenne, and A. R. Zakirov, Static magnetic susceptibility, crystal field and exchange interactions in rare earth titanate pyrochlores, *J. Phys.: Condens. Matter* 22, 276003 (2010)
36. B. K. Cho, B. N. Harmon, D. C. Johnston, and P. C. Canfield, Crystalline electric-field effects in single-crystal HoNi$_2$B$_2$C, *Phys. Rev.* B 53, 2217 (1996)
37. K. W. Chen, Y. Lai, Y. C. Chiu, S. Steven, T. Besara, D. Graf, T. Siegrist, T. E. Albrecht-Schmitt, L. Balicas, and R. E. Baumbach, Possible devil's staircase in the Kondo lattice CeSbSe, *Phys. Rev.* B 96, 014421 (2017)
38. S. V. Gallego, E. S. Tasci, G. de la Flor, J. M. Perez-Mato and M. I. Aroyo, Magnetic symmetry in the Bilbao Crystallographic Server: a computer program to provide systematic absences of magnetic neutron diffraction, *J. Appl. Cryst.* 45(6), 1236-1247 (2012)
39. H. Watanabe, H. C. Po, and A. Vishwanath, Structure and topology of band structures in the 1651 magnetic space groups, *Sci. Adv.* 4.8 eaat8685 (2018)
40. Z. D. Song, T. T. Zhang, Z. Fang, and C. Fang, Quantitative mappings between symmetry and topology in solids, *Nat. Commun.* 9, 3530 (2018)